\documentclass[nonacm,camera]{acmart}
\AtBeginDocument{%
  }

\usepackage{hyperref}
\usepackage{listings}
\usepackage{multirow}

\setcopyright{acmlicensed}
\copyrightyear{2026}
\acmYear{2026}
\acmConference[ASSETS '26]{The 28th International ACM SIGACCESS
Conference on Computers and Accessibility (accepted preprint)}{October 25--28,
  2026}{Porto, Portugal}

\begin{document}

\title{Quantifying the Engagement Trap: Impact of Short-form Video Recommender Systems on Users with ADHD}


\author{Vedad Misirlic}
\authornote{Both authors contributed equally to this research.}
\email{vedad.misirlic@tugraz.at}
\orcid{0009-0002-6382-9329}
\affiliation{
  \institution{Graz University of Technology}
  \city{Graz}
  \country{Austria}
}

\author{Gregor Mayr}
\authornotemark[1]
\email{gregor.mayr@tugraz.at}
\orcid{0009-0006-2836-6313}
\affiliation{%
  \institution{Graz University of Technology}
  \city{Graz}
  \country{Austria}
}
\author{Elisabeth Lex}
\affiliation{%
  \institution{Graz University of Technology}
  \city{Graz}
  \country{Austria}}
\orcid{0000-0001-5293-2967}
\email{elisabeth.lex@tugraz.at}

\renewcommand{\shortauthors}{Misirlic et al.}

\begin{abstract}
    Short-form video platforms use recommender systems to maximize engagement through highly efficient personalized recommendations. However, the impact of these recommendations on users with ADHD compared to users without ADHD remains underexplored. 
    Through this study, we introduce and operationalize the \emph{Engagement Trap}, illustrating how recommender systems, while successfully optimizing for engagement, disproportionately disadvantage users with ADHD. This stratified study of 302 participants, recruited via the online platform Prolific, compares experiences between participants with and without ADHD. Our results show that while recommendations are perceived as relevant across groups, participants with ADHD report significantly higher levels of time blindness, post-usage regret, and emotional distress when consuming recommendations. Moreover, we collect feedback for several proof-of-concept, theoretical design interventions for neuro-inclusive design principles. These findings provide quantitative evidence of systemic differences in engagement-optimized recommender systems and highlight the unbalanced negative effects and interactions these systems create for participants with ADHD. We argue for neurodiversity-aware, human-centered design approaches that mitigate such algorithmic harms and support more equitable experiences.
\end{abstract}



\settopmatter{printacmref=false} 
\setcopyright{none} 
\renewcommand\footnotetextcopyrightpermission[1]{} 
\pagestyle{plain}

\begin{abstract}
    Short-form video platforms use recommender systems to maximize engagement through highly efficient personalized recommendations. However, the impact of these recommendations on users with ADHD compared to users without ADHD remains underexplored. 
    Through this study, we introduce and operationalize the \emph{Engagement Trap}, illustrating how recommender systems, while successfully optimizing for engagement, disproportionately disadvantage users with ADHD. This stratified study of 302 participants, recruited via the online platform Prolific, compares experiences between participants with and without ADHD. Our results show that while recommendations are perceived as relevant across groups, participants with ADHD report significantly higher levels of time blindness, post-usage regret, and emotional distress when consuming recommendations. Moreover, we collect feedback for several proof-of-concept, theoretical design interventions for neuro-inclusive design principles. These findings provide quantitative evidence of systemic differences in engagement-optimized recommender systems and highlight the unbalanced negative effects and interactions these systems create for participants with ADHD. We argue for neurodiversity-aware, human-centered design approaches that mitigate such algorithmic harms and support more equitable experiences.
\end{abstract}
\maketitle

\section{Introduction}
\label{sec:intro}


Short-form video platforms, such as TikTok\footnote{\url{https://www.tiktok.com/}}, Instagram Reels\footnote{\url{https://www.instagram.com/reels/}}, and YouTube Shorts\footnote{\url{https://www.youtube.com/shorts}} have fundamentally transformed how people engage with digital content. These platforms are characterized by fast-paced short-form video recommendations, continuous playback, and an effectively infinite stream of content. Central to such platforms' experience are recommender systems that are typically optimized to maximize user engagement~\cite{su2021viewing, ye2022effects} by dynamically adapting recommendations based on interaction data such as watch time, re-watch rates, and interaction latency~\cite{chaney2018algorithmic}. As a result, users receive relevant and continuously novel content streams~\cite{zhang2021commentary} that can be difficult to disengage from~\cite{zhao2021analysis}.


Attention Deficit Hyperactivity Disorder (ADHD) is a neurodivergent condition commonly associated with differences in executive function, difficulties with impulse control, sustained attention, and time perception~\cite{faraone2021world,wender2001adults,tripp2009neurobiology}. 
While existing work focuses on understanding the challenges that neurodivergent individuals face in everyday life~\cite{carik2025exploring, khan2025inclusive}, to the best of our knowledge, there is a clear research gap in quantifying the negative effects of short-form video recommender systems on users with ADHD.

We address this gap and present a stratified quantitative study ($N=302$) that investigates self-reported short-form video consumption across three groups: participants with medically diagnosed ADHD, participants with self-reported ADHD, and participants without ADHD. We analyze participants' experiences related to disengagement challenges, time blindness (i.e. time agnosia), post-usage regret, and emotional impact. 
Our results show that participants with ADHD describe significantly greater difficulty disengaging, higher levels of time blindness , and increased post-usage regret and negative emotional impact. Notable, participants across all groups report that recommendations, in general, align with their interests and intent.
These findings indicate systematic differences in how short-form video recommendation systems impact users with and without ADHD. 



We term this phenomenon the \emph{Engagement Trap} --- an interaction model describing the interplay between: i) highly engaging content streams produced by recommender systems, and ii) the need for users to invest high efforts in actively disengaging from the content. Based on our findings, this \emph{Engagement Trap} affects users with ADHD significantly more negatively than users without ADHD.   

Building on these findings, we also explore several proof-of-concept intervention features on a theoretical basis and the participants' perception of these features.

Summing up, this work makes the following contributions:
\begin{itemize}
    \item The \emph{Engagement Trap}: an interaction model describing the interplay between engagement-optimized recommender systems delivering accurate and highly-personalized content, and significant disengagement efforts that affect users with ADHD more severely than users without ADHD.
    \item A stratified quantitative study ($N=302$) further operationalizing and validating the \emph{Engagement Trap} in self-reported short-form video recommendation consumption, and
    \item Empirical evidence of systematic differences in disengagement, time perception, and post-usage experiences between participants with and without ADHD.
\end{itemize}

\section{Related Work}
\label{sec:rw}
Understanding the \emph{Engagement Trap} requires examining both the underlying technologies that drive short-form video platforms and users' cognitive and emotional vulnerabilities. Accordingly, this section reviews the existing literature at the intersection of algorithmic recommendation, neurodiversity, and digital well-being. To contextualize the \emph{Engagement Trap}, we first examine the technical and psychological mechanics driving short-form video recommender systems. We then explore the specific cognitive vulnerabilities associated with ADHD in digital environments. Finally, we situate our work within the broader discourse on dark patterns and post-usage regret to highlight the systemic challenges neurodivergent users face when navigating infinite-scroll interfaces.

\subsection{Mechanics of Short-Form Video Recommendations}
Recommender systems serve as the core engine of engagement by using complex algorithms to predict user preferences in real-time. Taking past interactions and the general behavior of users into account, these systems create a feedback loop of highly personalized content~\cite{tong2023navigating}. The objective is to maximize user engagement to keep users on the platform and foster habitual usage~\cite{jannach2016recommendations}. In the context of short-form videos, these algorithms prioritize ``explorative'' content to prevent satiation, ensuring that the stream of stimuli remains unpredictable yet rewarding~\cite{zhao2021analysis}. These platforms mix highly appealing with less appealing content, thereby sustaining user engagement through intermittent reinforcement while simultaneously stretching the available content~\cite{zhang2021commentary}. Recommender systems leverage these psychological strategies to trigger neural reward mechanisms, resulting in dopamine release and maximized engagement~\cite{su2021viewing, ye2022effects}.

\subsection{ADHD and Digital Interfaces}


Existing work suggests that individuals with ADHD may be more vulnerable to problematic interaction patterns in digital environments, especially when interacting with short-form video interfaces~\cite{kim2019relationship, hong2021relationships, lin2024understanding, xu2025separation}. This is largely due to the infinite scroll mechanism, which creates an environment that exploits problems with executive functioning (i.e., cognitive control) and ability to implement rules for stopping~\cite{rixen2023loop}.
Specifically, users with ADHD interacting with short-form video platforms may experience increased executive dysfunction, such as \textit{time blindness}~\cite{barkley1997sense, nejati2020time} and poor \textit{impulse control}, leading to increased \textit{frustration}~\cite{seymour2019frustration, seymour2017adhd}.


Khan et al.~\cite{khan2025inclusive} suggest that recommender systems may inadvertently exploit neurobiological vulnerabilities in users with ADHD, while Seaver~\cite{seaver2019captivating} describes them as sociotechnical systems that can ``trap'' users into sustained usage. Furthermore, Carik et al.~\cite{carik2025exploring} offer a taxonomy of these challenges faced by neurodivergent users in digital environments. Despite these qualitative insights, there is a significant lack of quantitative evidence validating these experiences at scale, with only a few recent studies beginning to establish a connection between short-form video recommendations and clinical symptoms of inattentiveness~\cite{xu2025separation, chiencharoenthanakij2025short}. 

\subsection{Dark Patterns}
``Dark patterns'' are a concept used in user interface designs to intentionally mislead or manipulate users into unintended actions or prevent the user from performing a certain action. Recent comparative work by Mildner et al.~\cite{mildner2025comparative} revealed that users with ADHD are more adept at recognizing and avoiding dark patterns on social media than neurotypical users. However, short-form video platforms have a fundamentally different approach to ``hook" the user. Rather than relying on deceptive visual elements found in social media platforms, these systems function as sociotechnical traps~\cite{seaver2019captivating} that exploit difficulties with active decision-making and executive function. 
Building upon Mildner et al.'s findings, we argue that while users with ADHD might successfully navigate traditional interface dark patterns on social media platforms, they are more susceptible to the overstimulating, continuous stream of personalized short-form videos. This distinction highlights a critical gap in the literature regarding algorithmic harm against users with ADHD, a concept we term the \emph{Engagement Trap}.

\section{Study Design}
\label{sec:study}

To examine how short-form video recommendations affect users with and without ADHD, we recruited 302 participants via the online platform Prolific\footnote{\url{https://www.prolific.com/}}, which has already been used for similar studies in the past~\cite{khan2025inclusive}. By running a G*Power analysis~\cite{faul2009gpower31} post-hoc, we receive a $0.99$ power for the participant size of 302. Based on the previous research by Mildner et al.~\cite{mildner2025comparative}, for ethical reasons, Prolific does not collect medical data and relies on individuals' self-reporting their ADHD status. We further made the distinction between participants with self-reported and medically diagnosed ADHD. We did not collect medical records, but rather relied on the accuracy and honesty of participants in reporting having medically or self-diagnosed ADHD. Previous literature is mostly heterogeneous, only differing between self-reported ADHD and no ADHD groups, and our study makes the further distinction based on indicating whether participants received a medical diagnosis~\cite{dekkers2022understanding}.

Participants (149 female, 149 male, 3 diverse) ranged in age from 18 to 73 years (Median = 33). Of these, 150 participants reported having ADHD, including 65 with medically diagnosed ADHD ($ADHD_m$) and 85 with self-reported ADHD ($ADHD_{sr}$), while 152 participants reported not having ADHD ($No~ADHD$). Participants disclosed this information themselves, and no medical records were collected. Data on medication status were also not collected. The in-platform quotas were a 150-152 split of participants with and without ADHD (self-reported) and a 151-151 split between sexes on Prolific. Please note that Prolific only offers quota sampling for binary gender. In our survey, we included an additional option to select ``diverse'' or ``prefer not to disclose'' for identifying participants' gender. The division between $ADHD_m$ and $ADHD_{sr}$ was performed through collecting demographic data within the survey.

The questionnaire was self-developed and checked by two psychologists for face-validity. The questionnaire consisted of 7-point Likert-scale (1 (very low) to 7 (very high)) questions and required approximately five minutes to complete (Median = 4.6 minutes). Each participant was compensated with 0.90£ for completing the study (an hourly rate of about 11£). The study was approved by a university's Institutional Review Board (IRB). 


The questionnaire covered demographic information, platform usage, as well as four thematic areas: (1) time blindness (time agnosia) and disengagement difficulty, (2) post-usage regret, (3) emotional responses (i.e., feelings of guilt and frustration), and (4) preferences for a set of inclusive design features for short-form video platforms. Items within each thematic area demonstrated good internal consistency, with Cronbach’s alpha values exceeding $\alpha = 0.78$ across all areas.

The results were analyzed with two distinct statistical models to check for significance, but also to provide robustness to the statistical models. 

Firstly, given the presence of three distinct participant groups, we performed Kruskal-Wallis H-tests for each questionnaire item to assess group differences in response distributions. When statistically significant differences between participant groups were observed, we conducted Dunn's post-hoc pairwise comparisons using the Holm-Bonferroni correction.  Effect sizes are reported using epsilon squared ($\epsilon^2$) for Kruskal-Wallis H-tests and Cliff's Delta ($\delta$) for Dunn's post-hoc tests.
Secondly, the medically diagnosed ADHD ($ADHD_m$) and self-reported ADHD ($ADHD_{sr}$) were grouped in a unified ADHD group ($ADHD$). We performed a Mann-Whitney U-test, followed by Rank-Biserial Correlation ($r$) for effect sizes. 

In Section~\ref{sec:results}, we report the results from the three distinct groups, as they provide more granularity. The tests with two unique groups were used to validate the results and check whether there are differences between the results. This was done to further make the statistical models more robust in the context of being aware of self-reporting limitations and the lack of ability to validate whether participants were medically diagnosed with ADHD.

The full questionnaire is provided in Appendix~\ref{app:quest}.

\section{Results}
\label{sec:results}

\begin{figure}
    \centering
    \includegraphics[width=0.75\linewidth]{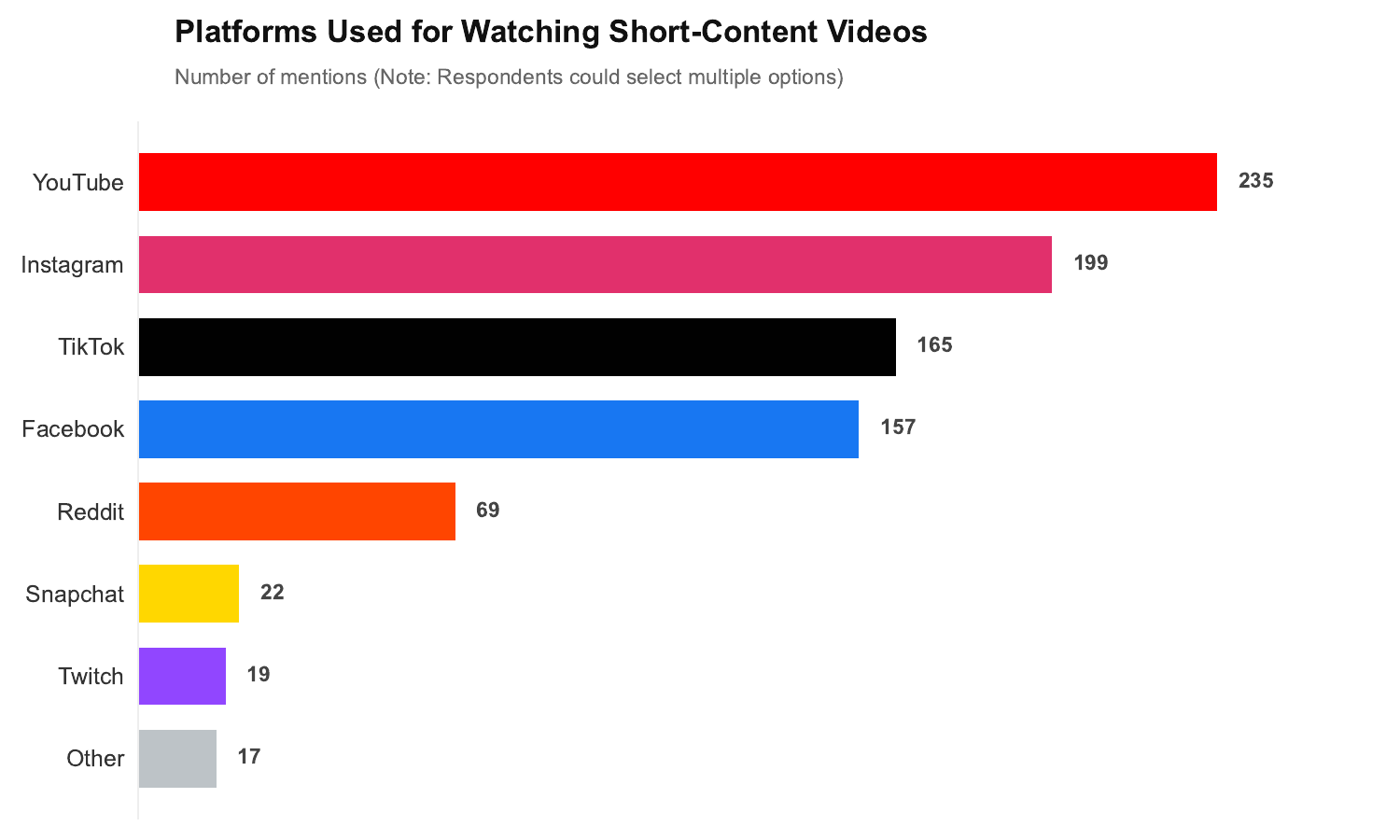}
    \caption{Self-reported platforms used for short-form video consumption. Participants were asked to select all platforms on which they regularly watch short-form content; therefore, the cumulative count exceeds the participation count. YouTube Shorts, Instagram Reels, and TikTok were the most frequently used platforms among the study cohort.}
    \Description{Self-reported platforms used for short-form video consumption. The Figure shows a bar chart outlining self-reported platforms usage. The most-used platform is YouTube Shorts ($75.9\%$), followed by Instagram Reels ($67.9\%$), and TikTok ($59.2\%$) }
    \label{fig:platform_use}
\end{figure}

We compare short-form video consumption experiences between participants with and without ADHD. As seen in Figure~\ref{fig:platform_use}, the most used platform is ``YouTube Shorts'' ($75.9\%$), followed by ``Instagram Reels'' ($67.9\%$), and TikTok ($59.2\%$). Participants were allowed to specify more than one platform, with the limitation that they could only specify platforms on which they watch short-form videos.

For the statistical tests reported over three distinct groups ($No~ADHD$, $ADHD_{sr}$, $ADHD_m$), Kruskal-Wallis tests revealed statistically significant overall group differences for all but one question. Dunn's post-hoc test showed no statistically significant difference between $ADHD_{sr}$ and $ADHD_m$. Hence, the $No~ADHD$ group serves as the reference group in post-hoc comparisons with Dunn's, revealing inter-group significances between the $No~ADHD$ and $ADHD_{sr}$ group, and the $No~ADHD$ and $ADHD_m$ group. Effect sizes vary between small and medium for statistically significant questions. Complete statistical results, including effects sizes, are reported in Appendix~\ref{app:KW_E2}.   

For the statistical tests reported over two distinct groups ($No~ADHD$, $ADHD$) Mann-Whitney U-test revealed statistically significant groups differences ($p < .01$) for all but one question. Effect sizes vary between small and medium for statistically significant question. Complete statistical results are reported in Appendix~\ref{app:MW_r}.

For simplicity, we report the statistical tests with three distinct groups, as they show a higher granularity of results. There were no significant differences in the results between the two-group and three-group comparison.

\begin{figure}[H]
    \centering
    \includegraphics[width=0.9\linewidth]{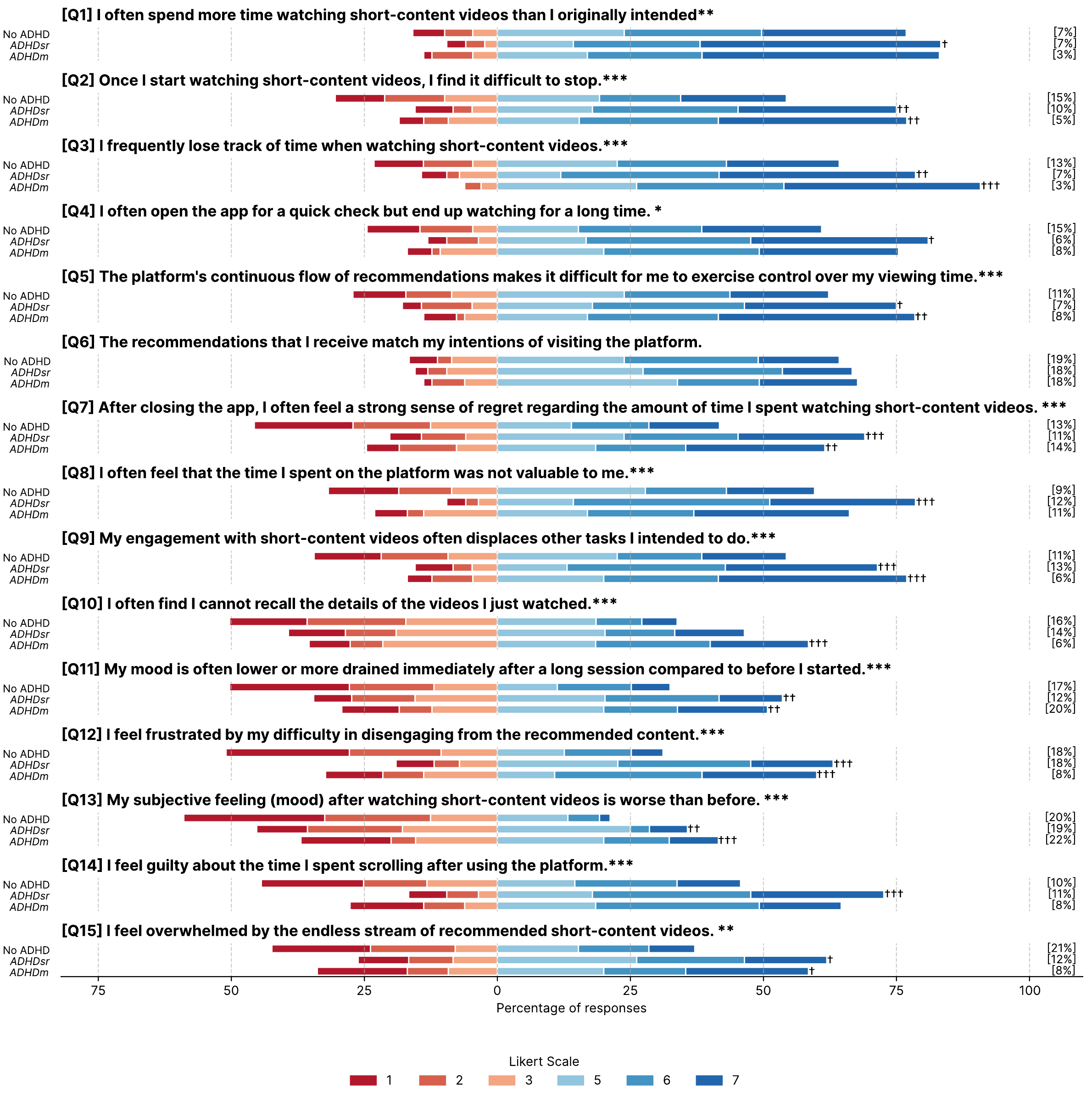}
    \caption{Survey responses to Likert questions (1 (very low) to 7 (very high)) measuring time blindness, frustration, feelings of guilt and post-usage regret in context of short-form video consumption. Respondents are grouped by ADHD status: medically diagnosed ADHD ($ADHD_m$), self-reported ADHD ($ADHD_{sr}$) or No ADHD ($No~ADHD$). Percentages shown in brackets represent the \% of respondents who were neutral in the response (i.e., Likert Score =  4). Asteriks denote Kruskall-Wallis statistical significances at $* = p < .05$, $** = p < .01$, and $*** = p < .001$. Daggers indicate statistical significances between the $No~ADHD$ group (control group) and $ADHD_{sr}$ and $ADHD_m$ based on Dunn's post-hoc tests: $\dag = p < .05$, $\dag\dag = p < .01$, and $\dag\dag\dag = p < .001$.}
    \label{fig:survey_results}
    \Description{A stacked bar chart showing Likert scale ratings (1 (very low) to 7 (very high)) for 15 questions (Q1–Q15) regarding short-content video recommendations. Responses are categorized into three groups: No ADHD, ADHDsr (self-reported), and ADHDm (medically diagnosed). Across almost all questions, the ADHD groups show significantly higher levels of agreement (blue bars, ratings 5–7) compared to the No ADHD group, particularly regarding difficulty stopping, losing track of time, and feeling regret or guilt after use. This is further emphasized through statistical significances between groups in almost all questions}
\end{figure}

Participants from all groups report that recommendations align with viewing intentions, with no statistical group difference observed ($p_{Q6}=.999$). Significant differences can be observed related to time blindness, feelings of guilt, frustration, and post-usage regret. Figure~\ref{fig:survey_results} shows response distributions for each question, along with statistical tests.



Participants with ADHD reported a significantly higher tendency to lose track of time while watching short-form videos compared to the $No~ADHD$ group ($p_{Q3} < .001$). As shown in Figure~\ref{fig:survey_results}, both $ADHD_{sr}$ and $ADHD_m$ groups are more likely to report watching longer than intended and experience difficulty stopping watching once started ($p_{Q2,Q5} < .001$). These findings indicate higher difficulty in disengaging for participants with ADHD.  This is often associated with the addictive nature of endless scrolling feeds.


Across all groups, participants reported difficulty disengaging from short-form video recommendations, with higher median difficulty reported by users with ($M (ADHD_m) = 5$, $M (ADHD_{sr}) = 5$) compared to users without ADHD ($M(No~ADHD) = 3$). Users with ADHD reported significantly higher frustration due to the difficulty in disengaging ($p_{Q9,Q12} < .001$) and higher levels of time blindness ($p_{Q1} = .008$), indicating difficulty perceiving how much time they spent consuming short-form video recommendations. 

The self-reported post-usage regret is significantly higher for both the $ADHD_{sr}$ and $ADHD_m$ groups ($p_{Q11, Q13, Q14} < .001$). These participants more frequently reported regretting the time spent on the platform, feeling that the time was not valuable, and that the usage displaced other intended activities. Their reported mood is significantly lower immediately after long sessions compared to their mood prior to engagement ($p_{Q11} < .001$). These experiences contribute to increased frustration and feelings of guilt. Participants with ADHD reported significantly greater frustration related to their difficulty in disengaging from the platforms ($p_{Q12} < .001$).



\begin{figure}[H]
    \centering
    \includegraphics[width=\linewidth]{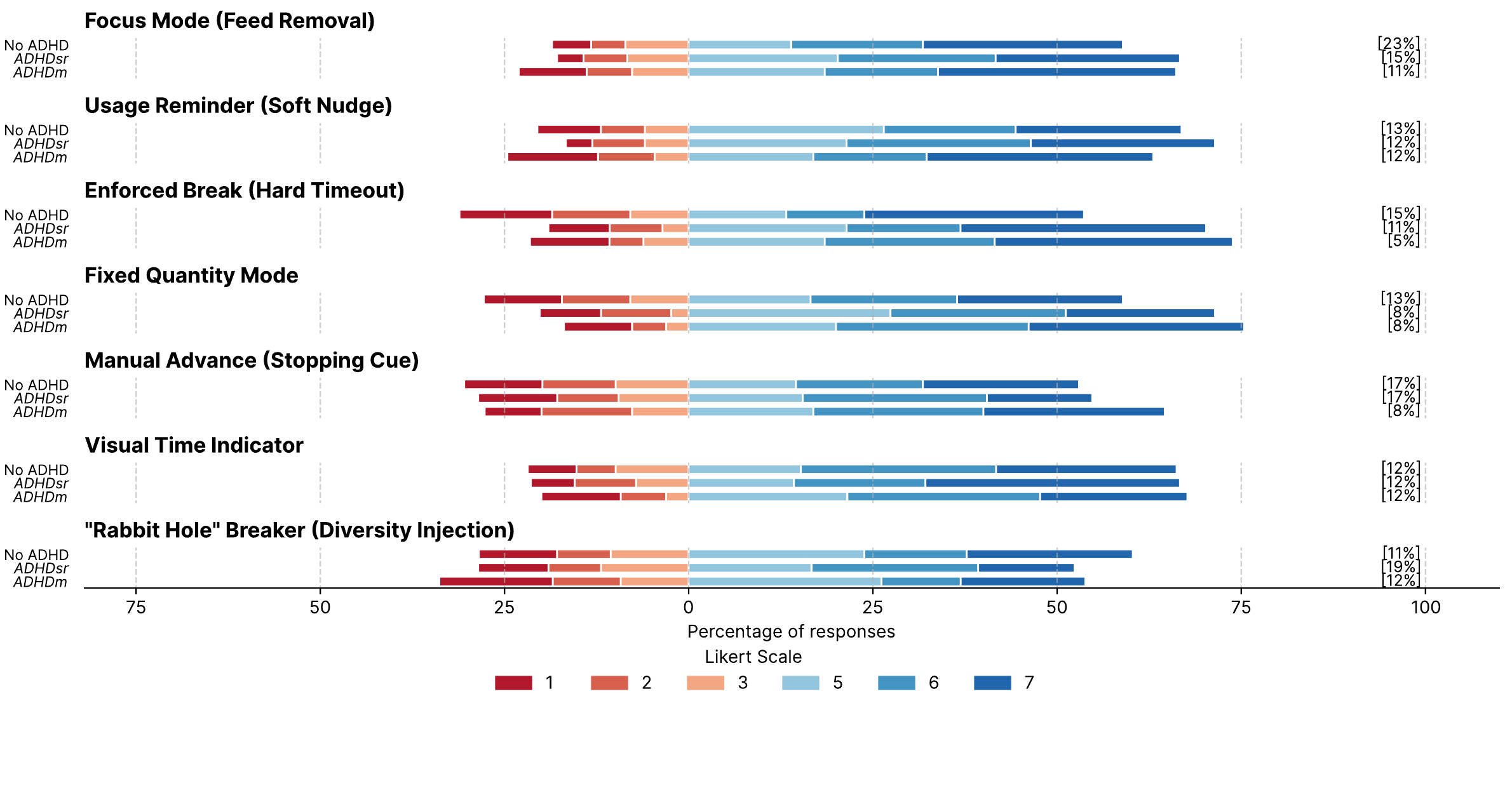}
    \caption{Survey results for the ``Tools'' subsection. Survey respondents answered questions about different inclusive features on a Likert scale (1 (very low) to 7 (very high)). Respondents are grouped into whether they have \textit{medically diagnosed ADHD} ($ADHD_m$), \textit{self-reported ADHD} ($ADHD_{sr}$), or \textit{No ADHD}. The percentages in the brackets on the right-hand side represent the \% of respondents who were neutral in the response (i.e., Likert-Score~=~4).}
    \label{fig:tools}
    \Description{A stacked bar chart showing Likert scale ratings (1 to 7) for 7 questions regarding potential inclusive features/tools. Responses are categorized into three groups: No ADHD, ADHDsr (self-reported), and ADHDm (medically diagnosed). Across almost all questions, the ADHD groups show higher levels of agreement (blue bars, ratings 5–7) compared to the No ADHD group.}
\end{figure}

We examined participant's interests in a set of suggested proof-of-concept, theorethical inclusive design features: (i) focus mode, (ii) usage reminder, (iii) enforced breaks, (iv) fixed quantity mode, (v) manual advance (removal of infinite scroll), (vi) a visual time indicator, and (vii) diversity injection. These design features were not implemented, but rather described in the questionnaire. Participants from all groups expressed positive attitudes toward these features (Median = 5), with \emph{fixed quantity mode}, and \emph{enforced break} receiving the highest median ratings. Full response distributions are visualized in Figure~\ref{fig:tools}.

\section{Discussion}
\label{sec:discussion}

Our findings highlight a systemic disparity in how individuals with and without ADHD experience short-form video recommendations. While algorithmic recommendations provided by various platforms seem to match the content that participants want to watch, individuals with ADHD experience greater difficulty disengaging and report higher levels of post-usage regret. As a result, the emotional cost of usage is significantly higher for individuals with ADHD. Moreover, the study results indicate that individuals with ADHD --- both self-reported and medically diagnosed --- suffer from the negative effects of short-form video recommendations significantly more than individuals without ADHD.

\subsection{Relevance Does Not Imply Control}
A central finding is the consensus between all groups that recommended content generally matches users' intentions when visiting the respective platforms and that the content is deemed appropriate to the reason of the platform visit. This suggests that the observed negative outcomes for individuals with ADHD are not driven by irrelevant or unwanted recommendations, but rather by how relevant content is delivered: since recommendations are so engaging, it is difficult to stop, in particular for individuals with ADHD. This finding points to a trade-off between recommending relevant, engaging content and supporting user control. This coincides with Khan et al.'s findings on inclusive recommender systems~\cite{khan2025inclusive}, which found that personalized recommendations foster a sense of community and social support in users, but simultaneously also impaired impulsivity and self-regulation. They argue that algorithms trap users by reinforcing immediate gratification over long-term goals.


\subsection{Disengagement, Time Perception, and Post-Usage Outcomes}

Participants with ADHD reported significantly higher levels of time blindness, disengagement difficulty, and post-usage regret compared to participants without ADHD.
These findings suggest a link between the \emph{Engagement Trap}, ``dark patterns'', and challenges in executive functioning: when no external stopping signals are available, users have to rely on self-regulation to intentionally disengage. This reliance may be particularly challenging for individuals with ADHD, for whom executive control and time management are known to require greater cognitive effort~\cite{weissenberger2021time,desrochers2019evaluation,brown2005attention}. 

The recommender system shifts the user's action from meaningful interactions to meaningless use, also called ``doomscrolling"~\cite{meinhardt2025scrolling}, as its intended function is to keep the user engaged to spend as much time on the platform as possible. Our results further reinforce this claim by showing significant differences in the post-usage regret, as well as very high levels of disengagement difficulties.

While frictionless design patterns like ``infinite scroll'' and ``autoplay''~\cite{rixen2023loop} push many users into a state of passive ``doomscrolling''~\cite{meinhardt2025scrolling}, our findings indicate that these mechanisms trigger a more profound cognitive trap for individuals with ADHD. Rather than experiencing the positive, intentional mental state of ``Flow'' --- where a person is fully and consciously involved in an activity~\cite{csikszentmihalyi1988flow} --- these users are pushed into a state of ``hyperfocus". This intense, extended state of attention is notoriously difficult to break without external cues, frequently resulting in negative consequences rather than productive engagement~\cite{sharpe2026influence}. Consequently, the \emph{Engagement Trap} exploits this specific neurodivergent vulnerability, locking users in until they can muster the significant cognitive effort required to manually disengage.

\subsection{Post-Usage Regret and Lowered Mood} 
Participants with ADHD reported higher levels of post-usage regret, including lower mood following consumption of short-content video recommendations compared to their mood prior to engagement. 
This regret and associated lowered mood arise from a conflicting goal: psychological friction that occurs when a user's behavior (scrolling for several hours) directly conflicts with their real-world goals (wanting to sleep or study).
The negative effects are likely compounded by the intense state of hyperfocus~\cite{hupfeld2019living}, which can cause neurodivergent users to lose track of time and exacerbate the resulting conflict.
These findings suggest that prolonged consumption of short-content video recommendations may be associated with increasingly negative outcomes for individuals with ADHD compared to neurotypical individuals.

\subsection{The Engagement Trap as an Interaction Model} 
Participants reported that short-form video recommendations --- although perceived as relevant recommendations --- affect their perception of time, with participants with ADHD reporting a statistically significant greater difficulty perceiving how much time they spend consuming short-form videos. The findings further suggest that individuals with ADHD have significantly greater difficulties in disengaging from the content, often feeling trapped or ``stuck in a loop'', similar to the ``dark patterns'' described by Mildner et al.~\cite{mildner2025comparative}. To conceptualize these dynamics, we introduce the \emph{Engagement Trap} as a condition in which (i) recommender systems deliver highly relevant content streams, and (ii) disengagement depends primarily on the internal self-regulation abilities of users or external triggers.

This interaction reflects a critical mechanism underlying the observed differences. When disengagement relies on self-regulation or external triggers, users with ADHD report greater difficulty perceiving time and disengaging intentionally once consumption has started.

Our findings suggest that current interaction patterns are associated with disproportionate challenges for users with ADHD, reflecting a mismatch between system design and cognitive characteristics.   

\subsection{Inclusive Features and Tools} 
A careful approach is needed to mitigate the \emph{Engagement Trap}, as overly restrictive interventions may be perceived as a threat to the users' autonomy~\cite{meinhardt2025scrolling}. Prior work has shown that interventions such as disabling algorithmic personalization and forced notifications~\cite{khan2025inclusive}, as well as small intentional design frictions~\cite{cox2016design}, can support disengagement while maintaining a sense of control.
In our study, we evaluated participants' perspectives on a range of potential design features on a conceptual level. Participants expressed consistently positive attitudes toward a range of features, including a focus mode, which removes the recommendation feed, usage reminders as a soft nudge, enforced breaks after a certain amount of time or videos passed, an option to only continue watching after pressing a button, a visual time indicator, and an intervention in the algorithmic recommendations. Our findings suggest that these features can function as an externalized control mechanism. Specifically, they can (i) introduce external stopping signals, reducing reliance on self-regulation,  (ii) limit time spent consuming content, and (iii) foster time awareness during consumption. Together, these mechanisms can address the \emph{Engagement Trap}, suggesting that accessibility in recommender systems may require more interface-level options that complement users' control over their engagement.



\subsection{Public Policy Implications}

Our results relate to ongoing public policy discussions. The ``European Accessibility Act (EAA)''\footnote{\url{https://commission.europa.eu/strategy-and-policy/policies/justice-and-fundamental-rights/disability/european-accessibility-act-eaa_en}} outlines accessibility requirements for digital products but does not explicitly address algorithmic transparency or the potential for engagement-driven design patterns to create disparities across user groups. Similarly, the ``Americans with Disabilities Act (ADA)''\footnote{\url{https://www.ada.gov/}} also applies to digital environments and advocates for more accessible features; however, it does not yet account for the ways in which algorithmic systems may impact neurodivergent users. 

Our findings suggest a need to extend existing regulatory frameworks to consider not only access to digital systems but also how interaction dynamics and optimization objectives influence outcomes across populations. Addressing such disparities is, however, critical to respect and protect the rights of neurodivergent individuals and to create a more inclusive, accessible, and transparent environment for all online users.


%

\section{Limitations \& Future Work}
\label{sec:limitations_fw}

This study has several limitations. First, ADHD status relied on participant disclosure. While we distinguish between medically diagnosed and self-reported participants, the lack of clinical verification may introduce potential variability of symptom severity within groups. 
However, the statistical alignment between the self-reported and diagnosed groups suggests the trends observed are robust across the spectrum of ADHD; only in some questions (Q4, Q8, Q10, Q14) do we observe a divergence between results in the medical and self-reported ADHD groups.

Second, the study is based on retrospective self-reported study data. Participants' perception of time and emotional state are subjective; for instance, a participant with ADHD might report that they lose track of time more often due to higher sensitivity to guilt (e.g., if they experience a social media addiction problem, it might happen more often, therefore, sensitivity might rise). Future work could complement self-reported measures with behavioral data, such as measuring the difference between perceived and actual loss of time (or usage/watch time).

Third, we did not collect information regarding ADHD medication intake, which might be an influencing factor in the statistical models. ADHD medication, which is found to temporarily restore emotional baselines, can also affect the results in whether the participants fall into the \emph{Engagement Trap}. As we did not collect data on medical intake, we cannot isolate whether ADHD medication served as a moderating factor in participants' interactions with the platforms or their susceptibility to the \emph{Engagement Trap}. Future longitudinal studies should also collect this data to account for potential moderating effects and to distinguish between medicated and unmedicated individuals. 

Fourth, the Likert-scale assessment of proof-of-concept inclusive features/tools (e.g., Focus Mode, Visual Time Indicator) was hypothetical. Participants rated these features based on descriptions rather than actual interaction. Future work should translate these inclusive design concepts into working applications to assess their impact on actual usage behavior and disengagement.

Ultimately, future work should explore longitudinal studies to capture the stronger effects of algorithmic harms and engagement traps. Neuro-inclusive design features are also of imperative importance in accessible design, hence, future research should focus on exploring these options further and creating accessible features and platforms that are built to offer fair and transparent algorithms, built for all individuals equally. 



\section{Conclusion}
\label{sec:conclusion}

This paper provides quantitative evidence that short-form video recommender systems affect participants with and without ADHD in fundamentally different ways. Based on survey data from 302 participants, we show that although engagement-optimized recommendations successfully deliver content aligned with viewing intentions, they systematically disadvantage participants with ADHD.

Participants with ADHD reported greater difficulty in disengaging, higher levels of time blindness, and experienced more negative emotional outcomes following short-form video consumption. We term these effects as the \emph{Engagement Trap}, an interaction model in which recommender systems optimized for engagement sustain attention in ways that undermine intentional disengagement and time perception.

Our findings demonstrate that engagement-optimized recommender systems impose a disproportionate emotional and behavioral burden on individuals with ADHD. Addressing this imbalance requires moving beyond engagement-centric optimization toward neurodiversity-aware, human-centered recommender system design and evaluation practices that explicitly support time awareness, intentional disengagement, and user control for individuals with ADHD.



\section{Declaration of Generative AI and AI-assisted Technologies in Writing}
\label{sec:ai}




During the preparation of this work, the author(s) used LanguageTool, Grammarly, and Gemini to perform grammar and spelling checks, improve writing style, paraphrase and reword, and format tables. After using these tools, the author(s) reviewed and edited the content as needed and take(s) full responsibility for the publication's content.

\bibliographystyle{ACM-Reference-Format}
\bibliography{references}
\newpage
 \appendix
 \clearpage
 \pagenumbering{roman}
 \setcounter{page}{1}
 \section{Questionnaire}
\label{app:quest}

\def\lstlistingname{Prompt}
\lstset{
  basicstyle=\ttfamily\footnotesize,
  columns=fullflexible,
  frame=single,
  breaklines=true,
  postbreak=\mbox{\textcolor{red}{$\hookrightarrow$}\space},
  breakatwhitespace=true,
  captionpos=b,
}

\begin{lstlisting}

Demographics:
    What is your age?
    What is your gender?
    What is your AD(H)D diagnosis?
    Which platforms do you use for watching short-content videos?
    If you marked "Other", please specify any other platforms you use.
    How many hours per day do you spend looking at short-content videos?

Time Blindness and Disengagement Difficulty:
    1. I often spend more time watching short-content videos than I originally intended.
    2. Once I start watching short-content videos, I find it difficult to stop.
    3. I frequently lose track of time when watching short-content videos.
    4. I often open the app for a quick check but end up watching for a long time. 
    5. The platform's continuous flow of recommendations makes it difficult for me to exercise control over my viewing time.
    6. The recommendations that I receive match my intentions of visiting the platform (e.g., educational content, general entertainment, etc.).

Post-usage regret:
    7. After closing the app, I often feel a strong sense of regret regarding the amount of time I spent watching short-content videos. 
    8. I often feel that the time I spent on the platform was not valuable to me.
    9. My engagement with short-content videos often displaces other tasks I intended to do.
    10. I often find I cannot recall the details of the videos I just watched.
    11. My mood is often lower or more drained immediately after a long session compared to before I started.

Feelings of Guilt and Frustration:
    12. I feel frustrated by my difficulty in disengaging from the recommended content.
    [Sanity Check] If you are an artificial model or an AI-assisted Agent, please answer with "five". Else, answer with "two".
    13. My subjective feeling (mood) after watching short-content videos is worse than before. 
    14. I feel guilty about the time I spent scrolling after using the platform.
    15. I feel overwhelmed by the endless stream of recommended short-content videos. 

Demand for Inclusive Features:
    Focus Mode (Feed Removal) 
    Usage Reminder (Soft Nudge)
    Enforced Break (Hard Timeout)
    Fixed Quantity Mode
    Manual Advance (Stopping Cue)
    Visual Time Indicator 
    "Rabbit Hole" Breaker (Diversity Injection)

\end{lstlisting}

\newpage

\section{Kruskal-Wallis and Epsilon-Squared Effect Size per Question}
\label{app:KW_E2}

\noindent

Table~\ref{tab:KW_E2} shows the Kruskal-Wallis p-values per question, as well as the $\epsilon^2$ test results. 

Table~\ref{tab:Dunn_Cliff} shows Dunn's Post Hoc p-values per question between groups, as well as Cliff's Delta 

\begin{table}[H]
    \centering
    \begin{tabular}{llrl}
\toprule
QID & p\_value & $\epsilon^2$ & $\epsilon\_label$ \\
\midrule
Q1 & 0.008 & 0.026 & small \\
Q2 & <0.001 & 0.048 & small \\
Q3 & <0.001 & 0.061 & medium \\
Q4 & 0.008 & 0.026 & small \\
Q5 & 0.001 & 0.040 & small \\
Q6 & 0.997 & -0.007 & negligible \\
Q7 & <0.001 & 0.060 & small \\
Q8 & <0.001 & 0.059 & small \\
Q9 & <0.001 & 0.064 & medium \\
Q10 & <0.001 & 0.043 & small \\
Q11 & <0.001 & 0.046 & small \\
Q12 & <0.001 & 0.108 & medium \\
Q13 & <0.001 & 0.050 & small \\
Q14 & <0.001 & 0.058 & small \\
Q15 & 0.001 & 0.037 & small \\
\bottomrule
\end{tabular}
    \caption{Kruskall-Wallis Significance Test per Question and $\epsilon^2$ Test}
    \label{tab:KW_E2}
    \Description{A table showing the p-values and the epsilon squared values for the 15 main questions asked. There are statistical significances for everything but one question and the effect sizes range from small to medium.}
\end{table}

\begin{table}[]
    \centering
    \begin{tabular}{|l|l|c|c|l|}
\hline
QID & Comparison & Dunn's (p-adj) & Cliff's Delta & Interpretation \\ \hline
\multirow{2}{*}{Q1} & No ADHD vs Self-reported ADHD & 0.017 & -0.210 & small \\
 & No ADHD vs medical ADHD & 0.046 & -0.187 & small \\ \hline
\multirow{2}{*}{Q2} & No ADHD vs Self-reported ADHD & 0.004 & -0.245 & small \\
 & No ADHD vs medical ADHD & 0.002 & -0.287 & small \\ \hline
\multirow{2}{*}{Q3} & No ADHD vs Self-reported ADHD & < 0.001 & -0.263 & small \\
 & No ADHD vs medical ADHD & < 0.001 & -0.325 & small \\ \hline
\multirow{2}{*}{Q4} & No ADHD vs Self-reported ADHD & 0.007 & -0.230 & small \\
 & No ADHD vs medical ADHD & 0.200 & -0.142 & negligible \\ \hline
\multirow{2}{*}{Q5} & No ADHD vs Self-reported ADHD & 0.018 & -0.203 & small \\
 & No ADHD vs medical ADHD & 0.002 & -0.279 & small \\ \hline
\multirow{2}{*}{Q6} & No ADHD vs Self-reported ADHD & 1.000 & -0.004 & negligible \\
 & No ADHD vs medical ADHD & 1.000 & -0.007 & negligible \\ \hline
\multirow{2}{*}{Q7} & No ADHD vs Self-reported ADHD & < 0.001 & -0.308 & small \\
 & No ADHD vs medical ADHD & 0.003 & -0.270 & small \\ \hline
\multirow{2}{*}{Q8} & No ADHD vs Self-reported ADHD & < 0.001 & -0.336 & medium \\
 & No ADHD vs medical ADHD & 0.034 & -0.197 & small \\ \hline
\multirow{2}{*}{Q9} & No ADHD vs Self-reported ADHD & < 0.001 & -0.284 & small \\
 & No ADHD vs medical ADHD & < 0.001 & -0.319 & small \\ \hline
\multirow{2}{*}{Q10} & No ADHD vs Self-reported ADHD & 0.045 & -0.179 & small \\
 & No ADHD vs medical ADHD & < 0.001 & -0.311 & small \\ \hline
\multirow{2}{*}{Q11} & No ADHD vs Self-reported ADHD & 0.003 & -0.259 & small \\
 & No ADHD vs medical ADHD & 0.004 & -0.263 & small \\ \hline
\multirow{2}{*}{Q12} & No ADHD vs Self-reported ADHD & < 0.001 & -0.405 & medium \\
 & No ADHD vs medical ADHD & < 0.001 & -0.358 & medium \\ \hline
\multirow{2}{*}{Q13} & No ADHD vs Self-reported ADHD & 0.005 & -0.242 & small \\
 & No ADHD vs medical ADHD & < 0.001 & -0.296 & small \\ \hline
\multirow{2}{*}{Q14} & No ADHD vs Self-reported ADHD & < 0.001 & -0.334 & medium \\
 & No ADHD vs medical ADHD & 0.055 & -0.187 & small \\ \hline
\multirow{2}{*}{Q15} & No ADHD vs Self-reported ADHD & 0.004 & -0.260 & small \\
 & No ADHD vs medical ADHD & 0.019 & -0.209 & small \\ \hline
\end{tabular}
    \caption{Dunn's Post Hoc Significance Test per Question and Cliff's Delta Test. Inter-group Cliff's Delta Test between $ADHD_{m}$ and $ADHD_{sr}$ was not performed, as there is no statistical significance on any of the questions for these two sub-groups.}
    \label{tab:Dunn_Cliff}
    \Description{A table showing Dunn's Test and Cliff's Alpha values for the 15 main questions asked. The results show statistical significances for all questions but one. The effect sizes are between small and medium, and the statistical significance is between the "no ADHD" and respectively self-reported and medical ADHD groups.}
\end{table}

\newpage

\section{Mann-Whitney U-test and Rank-Biserial Correlation per Question}
\label{app:MW_r}

Table~\ref{tab:MW_r} shows the Mann-Whitney U-test statistical significance results, and the Rank-Biserial Correlation ($r$) effect sizes per question.

\begin{table}[H]
    \centering
    \begin{tabular}{llcl}
\toprule
QID & p\_value & r & r\_label \\
\midrule
Q1 & <0.01 & 0.201 & small \\
Q2 & <0.001 & 0.263 & small \\
Q3 & <0.001 & 0.288 & small \\
Q4 & <0.01 & 0.189 & small \\
Q5 & <0.001 & 0.233 & small \\
Q6 & 1.00 & <0.01 & negligible \\
Q7 & <0.001 & 0.291 & small \\
Q8 & <0.001 & 0.283 & small \\
Q9 & <0.001 & 0.306 & medium \\
Q10 & <0.001 & 0.239 & small \\
Q11 & <0.001 & 0.262 & small \\
Q12 & <0.001 & 0.394 & medium \\
Q13 & <0.001 & 0.273 & small \\
Q14 & <0.001 & 0.283 & small \\
Q15 & <0.001 & 0.243 & small \\
\bottomrule
\bottomrule
\end{tabular}
    \caption{Mann-Whitney U-test and Rank-Biserial Correlation Test per Question}
    \label{tab:MW_r}
    \Description{A table showing the p-values and the r-values for the 15 main questions asked. There are statistical significances on everything but one question. The effect sizes vary between small and medium.}
\end{table}

\end{document}